\documentclass[a4paper]{article}
\usepackage{a4wide}
\usepackage[normalem]{ulem}
\usepackage{dsfont}
\usepackage[T1]{fontenc}
\usepackage{amsmath}
\usepackage{amssymb}
\usepackage{amsthm}
\usepackage[ansinew]{inputenc}
\usepackage{enumerate}
\usepackage{fancyhdr}
\usepackage{color}
\usepackage{ifpdf}
\usepackage{rotating}
\usepackage[T2A]{fontenc}
\usepackage{mathtext}

\DeclareMathOperator{\tr}{tr}

\title{Future non-linear stability for solutions of the Einstein-Vlasov system of Bianchi types II and VI$_0$}
\author{Ernesto Nungesser}

\begin{document}
   \newtheorem{deff}{Definition} 
    \theoremstyle{plain}
    \newtheorem{thm}{Theorem}
    \newtheorem{prop}{Proposition}
    \newtheorem{lem} {Lemma}
    \newtheorem{cor}{Corollary}
    \newcommand{\all}      {\quad \forall \ }
    \newcommand{\R}        {\mathds{R}}
    \newcommand{\N}        {\mathds{N}} 
\maketitle
\begin{abstract}
 In a recent paper \cite{E3} we have treated the future non-linear stability for reflection symmetric solutions of the Einstein-Vlasov system 
of Bianchi types II and VI$_0$. We have been able now to remove the reflection symmetry assumption, thus treating the non-diagonal case. Apart
from the increasing complexity the methods have been essentially the same as in the diagonal case, showing that they are thus quite powerful.
 Here the challenge was to put the equations in a form that permits the use of the previous results. We are able to conclude that after a possible
basis change the future of the non-diagonal spacetimes in consideration is asymptotically diagonal.
\end{abstract}

\section{The Einstein-Vlasov system}
A cosmological model represents a universe at a certain averaging scale. It is described via a Lorentzian metric $g_{\alpha\beta}$
 (we will use signature -- + + +) on a manifold $M$ and a family of fundamental observers. The metric is assumed to be time-orientable, 
which means that at each point of $M$ the two halves of the light cone can be labelled past and future in a way which varies continuously
 from point to point. This enables to distinguish between future-pointing and past-pointing timelike vectors. This is a physically
 reasonable assumption from both a macroscopic point of view e.g. the increase of entropy and also from a microscopic point of view e.g.
 the kaon decay.
One has also to specify the matter model and this we will do in the following.
 The interaction between the geometry and the matter is described by the Einstein field equations (we use geometrized units, i.e.
 the gravitational constant G and the speed of light in vacuum c are set equal to one):
\begin{eqnarray*}
G_{\alpha\beta}= 8\pi T_{\alpha \beta}
\end{eqnarray*}
where $G_{\alpha\beta}$ is the Einstein tensor and $T_{\alpha \beta}$ is the energy-momentum tensor. For the matter model
 we will take the point of view of kinetic theory \cite{St}. The sign conventions of \cite{RA} and the Einstein summation convention
 that repeated indices are to be summed over are used. Latin indices run from one to three and Greek ones from zero to three.

We will consider from now on that all the particles have \textit{equal} mass $m$. We will choose units such that $m=1$ which means that a distinction between velocities and momenta is not necessary.
 The collection of particles (galaxies or clusters of galaxies) will be described (statistically) by a non-negative real valued
 distribution function $f(x^\alpha,p^\alpha)$ on the mass shell. This function represents the density of particles at a given spacetime point with
 given four-momentum. Using the geodesic equations the restriction of the Liouville operator to the mass shell has the following form
\begin{eqnarray*}
 L=p^{\alpha}\frac{\partial}{\partial x^{\alpha}}-\Gamma^a_{\beta \gamma} p^{\beta} p^{\gamma}\frac{\partial}{\partial p^{a}}.
\end{eqnarray*}
where $\Gamma^a_{\beta \gamma}$ are the components of the metric connection.
We will consider the collisionless case which is described via the Vlasov equation:
\begin{eqnarray*}
L(f)=0
\end{eqnarray*}
The unknowns of our system are a 4-manifold $M$, a Lorentz metric $g_{\alpha\beta}$ on this manifold and the distribution function $f$
 on the mass shell defined by the metric. We have the Vlasov equation defined by the metric for the distribution function and the
 Einstein equations. It remains to define the energy-momentum tensor $T_{\alpha\beta}$ in terms of the distribution and the metric.
Before that we need a Lorentz invariant volume element on the mass shell. A point of a the tangent space has the volume element
$ |g^{(4)}|^{\frac{1}{2}} dp^0 dp^1 dp^2 dp^3$ ($g^{(4)}$ is the determinant of the spacetime metric) which is Lorentz invariant. Now considering $p^0$ as a dependent variable 
the induced (Riemannian) volume of the mass shell considered as a hypersurface in the tangent space at that point is
\begin{eqnarray*}
 \varpi=2H(p^{\alpha})\delta( p_{\alpha} p^{\alpha}+m^2)|g^{(4)}|^{\frac{1}{2}} dp^0 dp^1 dp^2 dp^3
\end{eqnarray*}
where $\delta$ is the Dirac distribution function and $H(p^{\alpha})$ is defined
 to be one if $p^{\alpha}$ is future directed and zero otherwise.
We can write this explicitly as:
 \begin{eqnarray*}
\varpi=|p_0|^{-1} |g^{(4)}|^{\frac{1}{2}} dp^1 dp^2 dp^3
\end{eqnarray*}
Now we define the energy momentum tensor as follows:
 \begin{eqnarray*}
T_{\alpha\beta}=\int f(x^{\alpha},p^{a}) p_{\alpha}p_{\beta}\varpi
\end{eqnarray*}
One can show that $T_{\alpha\beta}$ is divergence-free and thus it is compatible with the Einstein equations. For collisionless
 matter all the energy conditions hold. The Vlasov equation in a fixed spacetime can be solved by the method of characteristics:
\begin{eqnarray*}
\frac{dX^{a}}{ds}=P^{a}; \ \ \frac{dP^{a}}{ds}=-\Gamma^a_{\beta \gamma} P^{\beta} P^{\gamma}
\end{eqnarray*}
Let $X^a(s,x^{\alpha},p^a)$, $P^a(s,x^{\alpha},p^a)$ be the unique solution of that equation with initial conditions
 $X^a(t,x^{\alpha},p^a)=x^a$ and $P^a(t,x^{\alpha},p^a)=p^a$. Then the solution of the Vlasov equation can be written as:
\begin{eqnarray*}
f(x^{\alpha},p^a)=f_0(X^a(0,x^{\alpha},p^a),P^a(0,x^{\alpha},p^a))
\end{eqnarray*}
where $f_0$ is the restriction of $f$ to the hypersurface $t=0$. It follows that if $f_0$ is bounded the same is true for $f$. 
We will assume that $f$ has compact support in momentum space for each fixed t. This property
holds if the initial datum $f_0$ has compact support and if each hypersurface $t = t_0$ is a Cauchy hypersurface.
Before coming to our symmetry assumption we want to briefly introduce the initial value problem for the Einstein-Vlasov system.
 In general the initial data for the Einstein-matter equations consist of a metric $g_{ab}$ on the initial hypersurface, the second fundamental form $k_{ab}$ on that hypersurface and some matter data. Thus we have a Riemannian metric $g_{ab}$,
 a symmetric tensor $k_{ab}$ and some matter fields defined on an abstract 3-dimensional manifold $S$.

Solving the initial value problem means embedding $S$ into a 4-dimensional $M$ on which are defined a Lorentzian metric $g_{\alpha\beta}$
 and matter fields such that $g_{ab}$ and $k_{ab}$ are the pullbacks to $S$ of the induced metric and second fundamental form of the image of
 the embedding of $S$ while $f$ is the pullback of the matter fields. Finally $g^{\alpha\beta}$ and $f$ have to satisfy the Einstein-matter
 equations.

For the Einstein-Vlasov system it has been shown that given an initial data set there exists a corresponding solution
 of the Einstein-Vlasov system and that this solution is locally unique up to diffeomorphism. The extension
 to a global theorem has not been achieved yet. However if one assumes that the initial data have certain symmetry, this symmetry is inherited 
by the corresponding solutions. In particular for the case we will deal with, i.e. expanding Bianchi
 models (except type IX) coupled to dust or to collisionless matter the spacetime is future complete (theorem 2.1 of \cite{GC}).

\section{Bianchi spacetimes}
The basis for the classification of homogeneous spacetimes is the work of Bianchi which was introduced to cosmology by Taub. Here we will use the modern terminology and we define Bianchi spacetimes as follows:
\begin{deff}
A \textit{Bianchi spacetime} is defined to be a spatially homogeneous spacetime whose isometry group possesses a three-dimensional subgroup $G$
 that acts simply transitively on the spacelike orbits.
\end{deff}
Our results concern only a special class of the Bianchi spacetimes, namely that of class $A$.
\begin{deff}
A \textit{Bianchi A spacetime} is a Bianchi spacetime whose three-dimensional Lie algebra has traceless structure constants, i.e. $C^a_{ba}=0$.
\end{deff}
We will study II and VI$_0$. For Bianchi II the only non-vanishing structure constants are:
\begin{eqnarray}\label{sc2}
C^1_{23}=1=-C^1_{32}
\end{eqnarray}
and in the case of Bianchi VI$_0$ these are:
\begin{eqnarray}\label{sc6}
C^2_{31}=1=-C^2_{13},\ \ C^3_{21}=1=-C^3_{12}
\end{eqnarray}
We will use the metric approach. If $\mathbf{W}^a$ denote the 1-forms dual to the frame vectors $\mathbf{E}_a$
the metric of a Bianchi spacetime takes the form:
\begin{eqnarray}\label{mt}
^4 g=-dt^2+g_{ab}(t)\mathbf{W}^a \mathbf{W}^b
\end{eqnarray}
where $g_{ab}$ (and all other tensors) on $G$ will be described in terms of the frame components of a left invariant frame.
A dot above a letter will denote a derivative with respect to the cosmological time $t$.
We will use the 3+1 decomposition of the Einstein equations as made in \cite{RA}. Comparing our metric (\ref{mt}) with (2.28) of \cite{RA}
 we have that $\alpha=1$ and $\beta^a=0$ which means that the lapse function is the identity and the shift vector vanishes. There the abstract index 
notation is used. We can interpret the quantities as being frame components. There are different
 projections of the energy momentum tensor which are important 
\begin{eqnarray*}
   \rho&=&T^{00}\\
     j_a&=&T_a^0\\
  S_{ab}&=&T_{ab}
\end{eqnarray*}
where $\rho$ is the energy density and $j_a$ is the matter current.

The second fundamental form $k_{ab}$ can be expressed as:
\begin{eqnarray}\label{a}
 \dot{g}_{ab}=-2 k_{ab}.
\end{eqnarray}
The Einstein equations:
\begin{eqnarray}\label{EE}
 \dot{k}_{ab}=R_{ab}+k~k_{ab}-2k_{ac}k^c_b-8\pi(S_{ab}-\frac{1}{2}g_{ab}S)-4\pi\rho g_{ab}
\end{eqnarray}
where we have used the notations $S =g^{ab}S_{ab}$, $k = g^{ab}k_{ab}$, and $R_{ab}$ is the Ricci tensor of the three-dimensional metric.
The evolution equation for the mixed version of the second fundamental form is (2.35) of \cite{RA}:
\begin{eqnarray}\label{MV}
 {{\dot{k}}^a}_{b}=R^a_b+k~k^a_b - 8 \pi S^a_b + 4 \pi \delta^a_b (S -\rho)
\end{eqnarray}
From the constraint equations since $k$ only depends on the time variable we have that:
\begin{eqnarray}\label{CE1}
 R-k_{ab}k^{ab}+k^2&=&16\pi\rho\\
\label{CE2}
 \nabla^a k_{ab}& = & 8 \pi j_b
\end{eqnarray}
where $R$ is the Ricci scalar curvature.

Another useful relation concerns the determinant $g$ of the induced metric ((2.30) of \cite{RA}):
\begin{eqnarray}\label{det}
\frac{d}{dt}(\log g)=-2 k
\end{eqnarray}
Taking the trace of (\ref{MV}):
\begin{eqnarray}\label{im}
 \dot k=R+k^2+4\pi  S -12\pi \rho
\end{eqnarray}
With (\ref{CE1}) one can eliminate the energy density and (\ref{im}) reads:
\begin{eqnarray}\label{in}
 \dot k=\frac{1}{4}(k^2+R+3k_{ab}k^{ab})+4\pi S
\end{eqnarray}
Finally if one substitutes for the Ricci scalar with (\ref{CE1}):
\begin{eqnarray}\label{loo}
 \dot k=k_{ab}k^{ab}+4\pi (S+\rho)
\end{eqnarray}
Now with the 3+1 formulation our initial data are $(g_{ij}(t_0), k_{ij}(t_0),f(t_0))$, i.e. a Riemannian metric, a second fundamental
 form and the distribution function of the Vlasov equation, respectively, on a three-dimensional manifold $S(t_0)$. This is the initial data set
 at $t=t_0$ for the Einstein-Vlasov system.

We assume that $k < 0$ for all time following \cite{CC} (see comments below lemma 2.2 of \cite{CC}). This enables us to set without
 loss of generality $t_0=-2/k(t_0)$. The reason for this choice will become clear later and is of technical nature.

We will now introduce several new variables in order to use the ones which are common in Bianchi cosmologies and to be able
 to compare results. We can decompose the second fundamental form introducing $\sigma_{ab}$ as the trace-free part:
\begin{eqnarray}\label{TF}
 k_{ab}=\sigma_{ab}-H g_{ab}
\end{eqnarray}
\begin{eqnarray}\label{tf}
 k_{ab}k^{ab}=\sigma_{ab}\sigma^{ab}+3H^2
\end{eqnarray}
Using the Hubble parameter:
\begin{eqnarray*}
H=-\frac{1}{3}k
\end{eqnarray*}
we define:
\begin{eqnarray}\label{ab}
\Sigma_a^b=\frac{\sigma_a^b}{H}
\end{eqnarray}
and
\begin{eqnarray}
&&\Sigma_{+}=-\frac12(\Sigma_{2}^2+\Sigma_{3}^3) 
\\
&& \Sigma_{-}=-\frac{1}{2\sqrt{3}}(\Sigma_{2}^2-\Sigma_{3}^3) 
\end{eqnarray}
Thus
\begin{eqnarray*}
\Sigma_a^b=
\left(
\begin{matrix} 2\Sigma_+ & \Sigma^1_2 & \Sigma^1_3   \\
\Sigma^2_1  & -\Sigma_+-\sqrt{3}\Sigma_- & \Sigma^2_3 \\
\Sigma^3_1  & \Sigma^3_2  & -\Sigma_++\sqrt{3}\Sigma_-
\end{matrix} 
\right)
\end{eqnarray*}
The reason for using the variables $\Sigma_+$ and $\Sigma_-$ is that the diagonal case has been very important to understand the non-diagonal case.
Define also:
\begin{eqnarray}
\Omega=8\pi \rho/3H^2\\
\label{q} q=-1-\frac{\dot{H}}{H^2}\\
\label{tau} \frac{d\tau}{dt}=H
\end{eqnarray}
The time variable $\tau$ is dimensionless and sometimes very useful. From (\ref{CE1}) we obtain the constraint equation:
\begin{eqnarray*}
 \frac{1}{6H^2}(R-\sigma_{ab}\sigma^{ab})=\Omega-1
\end{eqnarray*}
and from (\ref{in}) the evolution equation for the Hubble variable:
\begin{eqnarray}\label{H-1}
\partial_t(H^{-1})=\frac{3}{2}+\frac{1}{12}(\frac{R}{H^2}+\frac{3}{H^2}\sigma_{ab}\sigma^{ab})+\frac{4\pi S}{3H^2}
\end{eqnarray}
Combining the last two equations with (\ref{MV}) we obtain the evolution equations for $\Sigma_-$ and $\Sigma_+$:
\begin{eqnarray}
\label{Bla1}&&\dot{\Sigma}_+=H[\frac{2R-3(R^2_2+R^3_3)}{6H^2}-\Sigma_+(3+\frac{\dot{H}}{H^2})+\frac{4\pi}{3H^2}(3S^2_2+3S^3_3-2S)]\\
\label{Bla2}&&\dot{\Sigma}_-=H[\frac{R_3^3-R^2_2}{2\sqrt{3}H^2}-(3+\frac{\dot{H}}{H^2})\Sigma_-+\frac{4\pi(S^2_2-S^3_3)}{\sqrt{3}H^2}]
\end{eqnarray}
Since we use a left-invariant frame $f$ will not depend on $x^a$ and the Vlasov equation takes the form:
\begin{eqnarray*}
p^0\frac{\partial f}{\partial t}-\Gamma^a_{\beta\gamma}p^{\beta}p^{\gamma}\frac{\partial f}{\partial p^a}=0
\end{eqnarray*}
It turns out that the equation simplifies if we express $f$ in terms of $p_i$ instead of $p^{i}$ what we can do due to the mass shell
 relation:
\begin{eqnarray*}
p^0\frac{\partial f}{\partial t}-\Gamma_{a\beta\gamma}p^{\beta}p^{\gamma}\frac{\partial f}{\partial p_a}=0
\end{eqnarray*}
Because of our special choice of frame the metric has the simple form (\ref{mt}). Due to the fact that we are contracting and
 the antisymmetry of the structure constants we finally arrive at:
\begin{eqnarray}\label{ve}
 \frac{\partial f}{\partial t}+(p^0)^{-1}C^d_{ba}p^{b}p_{d}\frac{\partial f}{\partial p_a}=0
\end{eqnarray}
From (\ref{ve}) it is also possible to define the characteristic curve $V_a$:
\begin{eqnarray}\label{charak}
 \frac{dV_a}{dt}=(V^0)^{-1}C^d_{ba}V^bV_{d}
\end{eqnarray}
for each $V_i(\bar{t})=\bar{v}_i$ given $\bar{t}$. Note that if we define:
\begin{eqnarray}\label{VVV}
 V=g^{ij}V_iV_j
\end{eqnarray}
due to the antisymmetry of the structure constants we have with (\ref{charak}):
\begin{eqnarray}\label{ha}
\frac{dV}{dt}=\frac{d}{dt}(g^{ij})V_iV_j
\end{eqnarray}
Let us also write down the components of the energy momentum tensor in our frame:
\begin{eqnarray}
&&T_{00}=\int f(t,p^{a}) p^0 \sqrt{g}dp^1 dp^2 dp^3\\
\label{emt2}&&T_{0j}=-\int f(t,p^{a}) p_j \sqrt{g}dp^1 dp^2 dp^3\\
\label{emt3}&&T_{ij}=\int f(t,p^{a}) p_i p_j (p^0)^{-1}\sqrt{g}dp^1 dp^2 dp^3
\end{eqnarray}

\section{The asymptotics of Bianchi II and VI$_0$}
Before coming to the non-diagonal case we have a look at the tilted fluid models, 
since they are non-diagonal as well and they may help us to understand the non-diagonal case with
collisionless matter. For the tilted Bianchi II we use the corresponding equations of \cite{HBW} 
and for Bianchi VI$_0$ the equations of \cite{Hervik}, in both cases with $\gamma=1$. We will not
go into the details for this we refer to the mentioned work. The point is that looking at the linearization
we see that the variables which did not appear in the diagonal case have decay rates which are between
the ones considered previously. This is a good sign. Also in \cite{BarrowH} the stability of the Ellis-MacCallum solution,
 in fact the stability of the Collins solution, was already considered within the Einstein-Euler system.

\subsection{Equations of the non-diagonal case}
Using (\ref{ab}) we arrive with (\ref{MV}) for $a\neq b$ to:
\begin{eqnarray*}
\dot{\Sigma}_a^b=H[\frac{R_a^b}{H^2}-\Sigma_a^b(3+\frac{\dot{H}}{H^2})-\frac{8\pi S^b_a}{H^2}]; \ \ a\neq b
\end{eqnarray*}
which together with (\ref{Bla1})-(\ref{Bla2}), i.e.
\begin{eqnarray*}
&&\dot{\Sigma}_+=H[\frac{2R-3(R^2_2+R^3_3)}{6H^2}-\Sigma_+(3+\frac{\dot{H}}{H^2})+\frac{4\pi}{3H^2}(3S^2_2+3S^3_3-2S)]\\
&&\dot{\Sigma}_-=H[\frac{R_3^3-R^2_2}{2\sqrt{3}H^2}-(3+\frac{\dot{H}}{H^2})\Sigma_-+\frac{4\pi(S^2_2-S^3_3)}{\sqrt{3}H^2}]
\end{eqnarray*}
 describe the evolution of $\Sigma^a_b$. The expression for the Ricci tensor is:
\begin{eqnarray}\label{ricci}
 R_{ij}=-\frac{1}{2}C^l_{ki}(C^k_{lj}+g_{lm}g^{kn}C^m_{nj})-\frac{1}{4}C^m_{nk}C^p_{ql}g_{jm}g_{ip}g^{kq}g^{ln}
\end{eqnarray}
and
\begin{eqnarray}\label{racci}
 R_i^j=R_{ib}g^{bj}=-\frac{1}{2}C^l_{ki}g^{bj}(C^k_{lb}+g_{lm}g^{kn}C^m_{nb})-\frac{1}{4}C^j_{nk}C^p_{ql}g_{ip}g^{kq}g^{ln}
\end{eqnarray}
We will now derive some expression concerning the derivative of (\ref{ricci}):
\begin{eqnarray*}
 \dot{R}_{ij}&=&C^l_{ki}C^m_{nj}(k_{lm}g^{kn}-g_{lm}k^{kn})+\\
& &\frac{1}{2}C^m_{nk}C^p_{ql}(k_{jm}g_{ip}g^{kq}g^{ln}+g_{jm}k_{ip}g^{kq}g^{ln}-g_{jm}g_{ip}k^{kq}g^{ln}-g_{jm}g_{ip}g^{kq}k^{ln})
\end{eqnarray*}
Thus:
\begin{eqnarray*}
 g^{jr}\dot{R}_{ij}&=& g^{jr}C^l_{ki}C^m_{nj}(k_{lm}g^{kn}-g_{lm}k^{kn})+\\
& &\frac{1}{2}C^p_{ql}[C^m_{nk}k^r_mg_{ip}g^{kq}g^{ln}+C^r_{nk}(k_{ip}g^{kq}g^{ln}-g_{ip}(k^{kq}g^{ln}+g^{kq}k^{ln}))]
\end{eqnarray*}
For $r=i$ and relabelling the $m$ with $i$ for the terms with the prefactor $\frac12$:
\begin{eqnarray*}
 g^{ji}\dot{R}_{ij}= g^{ji}C^l_{ki}C^m_{nj}(k_{lm}g^{kn}-g_{lm}k^{kn})+\frac{1}{2}C^p_{ql}C^i_{nk}[2k_{ip}g^{kq}g^{ln}-g_{ip}(k^{kq}g^{ln}+g^{kq}k^{ln}))]
\end{eqnarray*}
Rearranging terms:
\begin{eqnarray*}
 g^{ji}\dot{R}_{ij}= C^l_{ki}C^m_{nj}(k_{lm}g^{kn}g^{ji}-g_{lm}k^{kn}g^{ji})+C^p_{ql}C^i_{nk}[k_{ip}g^{kq}g^{ln}-g_{ip}k^{kq}g^{ln}].
\end{eqnarray*}
We see that the first with the third and the second with the fourth term cancel each other, hence:
\begin{eqnarray}\label{back}
 g^{ji}\dot{R}_{ij}=0
\end{eqnarray}
The evolution equation for the Ricci scalar due to (\ref{back}) is:
\begin{eqnarray*}
 \dot{R}=2R^i_jk^j_i=2H(-R+R^i_j\Sigma^j_i)
\end{eqnarray*}
Define
\begin{eqnarray*}
 N^j_i=\frac{R^j_i}{H^2}
\end{eqnarray*}
The derivative of this expression is:
\begin{eqnarray*}
\dot{N}^j_i=\frac{g^{pj}\dot{R}_{pi}}{H^2}+2H(N^p_i\Sigma^j_p-(1+\frac{\dot{H}}{H^2})N^j_i)
\end{eqnarray*}
Consider the quantity $N=R/H^2$. Its evolution equation is:
\begin{eqnarray}\label{NNN}
 \dot{N}=2H[q N+N^i_j\Sigma^j_i]
\end{eqnarray}
\subsection{Curvature expressions}
For bookkeeping reasons we define the following quantities where we use
from now on $g$ for the determinant of the metric.
\begin{eqnarray*}
&&A=g^{22}g^{33}-(g^{23})^2=\frac{g_{11}}{g}; \ \ \ B=g^{13}g^{23}-g^{12}g^{33}=\frac{g_{12}}{g}\\
&&C=g^{12}g^{23}-g^{13}g^{22}=\frac{g_{13}}{g}; \ \ \  D=g^{12}g^{13}-g^{11}g^{23}=\frac{g_{23}}{g}\\
&&E=g^{11}g^{33}-(g^{13})^2=\frac{g_{22}}{g}; \ \ \ F=g^{11}g^{22}-(g^{12})^2=\frac{g_{33}}{g}\\
\end{eqnarray*}
Let us denote the quantities divided by $H^2$ with small letters, i.e. $a=\frac{A}{H^2}$.
\subsubsection{Curvature expressions for Bianchi II}
Using (\ref{racci}) for Bianchi II:
\begin{eqnarray*}
 R_i^j=\frac{1}{2}g_{11}[C^1_{2i}(g^{23}g^{2j}-g^{22}g^{3j})+C^1_{i3}(g^{23}g^{3j}-g^{33}g^{2j})]
+\frac{1}{2}g_{i1}C^j_{23}A
\end{eqnarray*}
We obtain:
\begin{eqnarray*}
 R=-\frac{1}{2}g_{11}A=-\frac{1}{2}\frac{(g_{11})^2}{g}
\end{eqnarray*}
and as in the diagonal case:
\begin{eqnarray*}
R^1_1=-R=-R^2_2=-R^3_3\\
R_1^2=R_1^3=R_2^3=R^2_3=0
\end{eqnarray*}
However in the non-diagonal case we have:
\begin{eqnarray*}
R_2^1=-2\frac{g_{12}}{g_{11}}R\\
R_3^1=-2\frac{g_{13}}{g_{11}}R
\end{eqnarray*}
Thus
\begin{eqnarray*}
 \dot{N}=-2H[(1+\frac{\dot{H}}{H^2}+4\Sigma_+)N-W_{II}]
\end{eqnarray*}
where $W_{II}=N^1_2\Sigma^2_1+N^1_3\Sigma^3_1$. In order to calculate the derivative of $N^1_2$ we need the following expression:
\begin{eqnarray*}
 R\frac{d}{dt}(-2\frac{g_{12}}{g_{11}})=2H[2\Sigma^1_2R+(3\Sigma_++\sqrt{3}\Sigma_-)R^1_2-\frac{1}{2R} ((R^1_2)^2\Sigma^2_1+R^1_3R^1_2\Sigma^3_1)]
\end{eqnarray*}
Hence:
\begin{eqnarray*}
 \dot{N}^1_2=H[4N\Sigma^1_2-2(\Sigma_++1-\sqrt{3}\Sigma_-+\frac{\dot{H}}{H^2})N^1_2+W^1_2]\\
 \dot{N}^1_3=H[4N\Sigma^1_3-2(\Sigma_++1+\sqrt{3}\Sigma_-+\frac{\dot{H}}{H^2})N^1_3+W^1_3]
\end{eqnarray*}
where
\begin{eqnarray*}
 W^1_2=-2\Sigma^3_2N^1_3+ N^1_2N^{-1}(\Sigma^2_1N^1_2+\Sigma^3_1N^1_3)\\
W^1_3=-2\Sigma^2_3N^1_2+ N^1_3N^{-1}(\Sigma^2_1N^1_2+\Sigma^3_1N^1_3)
\end{eqnarray*}
\subsubsection{Curvature expressions for Bianchi VI$_0$}

With (\ref{racci}) we obtain:
\begin{eqnarray*}
 -2R_i^j=&&g^{1j}(C^3_{2i}+C^2_{3i})+g_{i2}(C^j_{13}E-C^j_{12}D)+g_{i3}(-C^j_{13}D+C^j_{12}F)\\
&&+g_{22}[C^2_{1i}(-g^{3j}g^{11}+g^{1j}g^{13})+C^2_{3i}(-g^{3j}g^{31}+g^{1j}g^{33})]\\
&&+g_{33}[C^3_{1i}(-g^{2j}g^{11}+g^{1j}g^{12})+C^3_{2i}(-g^{2j}g^{21}+g^{1j}g^{22})]\\
&&+g_{23}[C^2_{1i}(g^{1j}g^{12}-g^{2j}g^{11})+C^2_{3i}(g^{1j}g^{23}-g^{2j}g^{13})+C^3_{1i}(g^{1j}g^{13}-g^{3j}g^{11})+C^3_{2i}(g^{1j}g^{23}-g^{3j}g^{21})]
\end{eqnarray*}
In particular:
\begin{eqnarray*}
 &&R=-\frac12[(\sqrt{g_{22}E}+\sqrt{g_{33}F})^2-4g_{23}D]=-\frac{1}{2g}[(g_{22}+g_{33})^2-4g_{23}^2]\\
 &&R_2^2=\frac12(g_{22}E-g_{33}F)=\frac{1}{2g}[(g_{22})^2-(g_{33})^2]
\end{eqnarray*}
and like in the diagonal case:
\begin{eqnarray*}
&&R=R_1^1\\
&&R_2^2=-R^3_3\\
&&R_2^1=R_3^1=0
\end{eqnarray*}
However we have
\begin{eqnarray*}
&&N^3_2=-N^2_3=g_{23}(f-e)=-N_{23}(N_3+N_2)\\
&&N_1^2=-2\frac{g^{12}}{H^2}+g_{12}(e-f)=N_{12}(N_2-N_3)-2N_{13}N_{23}\\
&&N_1^3=-2\frac{g^{13}}{H^2}+g_{13}(f-e)=N_{13}(N_2-N_3)-2N_{12}N_{23}
\end{eqnarray*}
where $N_{ij}$ is defined as
\begin{eqnarray*}
 N_{ij}=\frac{g_{ij}}{\sqrt{g}H}
\end{eqnarray*}
and 
\begin{eqnarray*}
N_2&=&N_{22}\\
N_3&=&-N_{33}
\end{eqnarray*}
which means that $N^2_2=R^2_2/H^2$:
\begin{eqnarray*}
 N^2_2=\frac12((N_2)^2-(N_3)^2)
\end{eqnarray*}

Recalling that 
\begin{eqnarray*}
 \frac{\dot{g}}{g}=6H
\end{eqnarray*}
we can compute the derivatives of $N_{ij}$ using the following formula

\begin{eqnarray*}
 \dot{N}_{ij}=H[qN_{ij}-2\Sigma^l_iN_{lj}]
\end{eqnarray*}
Hence
 \begin{eqnarray}
\dot{N}_{12}=H[(q-4\Sigma_+)N_{12}-2\Sigma^2_1N_2-2\Sigma^3_1N_{23}]\\
\dot{N}_{13}=H[(q-4\Sigma_+)N_{13}-2\Sigma^2_1N_{23}+2\Sigma^3_1N_3]\\
\dot{N}_{23}=H[(2\Sigma_++2\sqrt{3}\Sigma_-+q)N_{23}+2\Sigma^3_2N_3-2\Sigma^1_2N_{13}]\\
\dot{N}_2=H[(2\Sigma_++2\sqrt{3}\Sigma_-+q)N_2-2\Sigma^1_2N_{12}-2\Sigma^3_2N_{23}]\\
\dot{N}_3=H[(2\Sigma_+-2\sqrt{3}\Sigma_-+q)N_3+2\Sigma^1_3N_{13}+2\Sigma^2_3N_{23}]
\end{eqnarray}
From (\ref{NNN}) we obtain
\begin{eqnarray}
 \dot{N}=2H[(2\Sigma_++q)N-2\sqrt{3}\Sigma_-N^2_2+N^2_3\Sigma^3_2+N^3_2\Sigma^2_3+N^2_1\Sigma^1_2+N^3_1\Sigma^1_3]
\end{eqnarray}
The evolution equation for $N^2_2$:
\begin{eqnarray}
 \dot{N}^2_2=H[2(2\Sigma_++q)N^2_2+2\sqrt{3}\Sigma_-((N_3)^2+(N_2)^2)-2(\Sigma^1_2N_{12}N_2+\Sigma^1_3N_{13}N_3+\Sigma^3_2N_{23}N_2+\Sigma^2_3N_{23}N_3)]
\end{eqnarray}

\subsection{The non-diagonal asymptotics of Bianchi II and VI$_0$}
We will now discuss the asymptotics of the non-diagonal case. The structure of the analysis is very similar to the diagonal case. 
We start with a bootstrap argument and end with applying Arzela-Ascoli. Next we will collect the bootstrap assumptions.
The prefactors denoted by $A$ and some index are small constants. 
\subsubsection{Bootstrap assumptions for Bianchi II}
\begin{eqnarray*}
|\Sigma_+-\frac{1}{8}| &\leq& A_+ (1+t)^{-\frac{3}{8}}\\
 |N+\frac{9}{32}| &\leq& A_c (1+t)^{-\frac{3}{8}}\\
|\Sigma^2_3|&\leq& A_{23} (1+t)^{-\frac{3}{8}}\\
|\Sigma^3_2|&\leq& A_{32} (1+t)^{-\frac{3}{8}}\\
|{\Sigma}^1_2|&\leq& A_{12}\\
|{\Sigma}^1_3|&\leq& A_{13}\\
|{N}_2^1|&\leq& A_{c12}\\
|{N}_3^1|&\leq& A_{c13}\\
P&\leq& A_m (1+t)^{-\frac{1}{3}}\\
|\Sigma_-|&\leq& A_-(1+t)^{-\frac{3}{4}}\\
|\Sigma^2_1|&\leq& A_{21} (1+t)^{-\frac{3}{4}}\\
|\Sigma^3_1|&\leq& A_{31} (1+t)^{-\frac{3}{4}}
\end{eqnarray*}
\subsubsection{Bootstrap assumptions for Bianchi VI$_0$}
\begin{eqnarray*}
|\Sigma_++\frac{1}{4}| &\leq& A_+ (1+t)^{-\frac{3}{8}}\\
|\Sigma_-|&\leq &A_-(1+t)^{-\frac{3}{8}}\\
|N+\frac98|& \leq &A_{c1} (1+t)^{-\frac{3}{8}}\\
|N^2_2| &\leq& A_{c2} (1+t)^{-\frac{3}{8}}\\
|N_{12}|&\leq& C_1 \\
|N_{13}|&\leq& C_2 \\
|N_{23}|&\leq& A_{c23}\\
P&\leq &A_m (1+t)^{-\frac{1}{3}}\\
|\Sigma^2_3|&\leq& A_{23} (1+t)^{-\frac{3}{4}}\\
|\Sigma^3_2|&\leq& A_{32} (1+t)^{-\frac{3}{4}}\\
|\Sigma^1_2|&\leq& A_{12} (1+t)^{-\frac{3}{4}}\\
|\Sigma^1_3|&\leq& A_{13} (1+t)^{-\frac{3}{4}}\\
|\Sigma^3_1|&\leq& C_3  \\
|\Sigma^2_1|&\leq& C_4  
\end{eqnarray*}
\subsubsection{Mean curvature}
Concerning the estimate of $H$ there is no difference with respect to
the diagonal case. The reason is that the estimate of $D$
\begin{eqnarray*}
D=\frac{1}{12}(N+\frac{3}{H^2}\sigma_{ab}\sigma^{ab})+\frac{4\pi S}{3H^2}
\end{eqnarray*}
is the same. Thus as in the diagonal case it follows from (\ref{H-1}) that
\begin{eqnarray*}
 \partial_t(H^{-1})=\frac32 + O(\epsilon t^{-\frac38})
\end{eqnarray*}
and following the steps made for the diagonal case we arrive at:
\begin{eqnarray*}
\boxed{H=\frac{2}{3}t^{-1}(1+O(\epsilon t^{-\frac{3}{8}}))}
\end{eqnarray*}
will hold.
\subsubsection{Estimate of the metric and P}
For a matrix $A$ its norm can be defined as:

\begin{eqnarray*}
\|A\|=\sup\{ |A x|/|x|: x\ne 0\}
\end{eqnarray*}

Let $B$ and $C$ be $n\times n$ symmetric matrices with $C$ positive definite. It is possible to define a \textit{relative norm} by:

\begin{eqnarray*}
\|B\|_C=\sup\{ |B x|/|C x|: x\ne 0\}
\end{eqnarray*}

Clearly:

\begin{eqnarray*}
\|B\| \le \|B\|_C \|C\|
\end{eqnarray*}

It also true that:

 \begin{eqnarray}\label{trick}
\|B\|_C \leq \sqrt{ \tr(C^{-1}BC^{-1}B)}
 \end{eqnarray}

This can be shown as follows. Consider the common eigenbasis $b_i$ of $B$ and $C$. Then there exist $\alpha_i$
 such that $B b_i= \alpha_i C b_i$ for each $i$. Then (\ref{trick}) is equivalent to the statement that the maximum modulus of any
 $\alpha_i$ is smaller than $\Sigma_i \alpha_i^2$. Using (\ref{trick}) we obtain in the sense of \textit{quadratic forms}:

\begin{eqnarray}\label{si}
\sigma^{ab}\leq(\sigma_{cd}\sigma^{cd})^{\frac{1}{2}}g^{ab}
\end{eqnarray}
Define
\begin{eqnarray*}
\bar{g}^{ab}=t^{\frac{p}{q}}g^{ab}
\end{eqnarray*}
Then
\begin{eqnarray*}
\frac{d}{dt}(t^{-\gamma}\bar{g}^{ab})= t^{-\gamma-1}\bar{g}^{ab}(-\gamma+\frac{p}{q})+2t^{-\gamma+\frac pq}(\sigma^{ab}-Hg^{ab})
\end{eqnarray*}
where we have introduced for technical reasons a small positive parameter $\gamma$. Using now the inequality (\ref{si})

\begin{eqnarray}\label{dec}
\frac{d}{dt}(t^{-\gamma}\bar{g}^{ab})\leq t^{-\gamma-1}\bar{g}^{ab}[-\gamma+\frac{p}{q}+2tH((H^{-2}\sigma_{cd}\sigma^{cd})^{\frac{1}{2}}-1)]
\end{eqnarray}
Using the equation (\ref{dec}) and the estimate of $H$
\begin{eqnarray*}
\frac{d}{dt}(t^{-\gamma}\bar{g}^{ab})\leq t^{-\gamma-1}\bar{g}^{ab}[-\gamma+\frac{p}{q}+\frac43(1+O(\epsilon t^{-\frac38}))((H^{-2}\sigma_{cd}\sigma^{cd})^{\frac{1}{2}}-1)]
\end{eqnarray*}
We obtain decay for the metric (in the sense of quadratic forms) provided that $(H^{-2}\sigma_{cd}\sigma^{cd})^{\frac{1}{2}}\leq1$. 
This holds for Bianchi II and VI$_0$ with for instance $\frac{p}{q}=0.4$. Thus we have
\begin{eqnarray*}
 g^{ab} \leq t^{-\frac{p}{q}} t_0^{\frac{p}{q}}g^{ab}(t_0)
\end{eqnarray*}
This implies that the components of the metric are also bounded by some constant $C(t_0)$ which depends on the terms of $g^{ab}(t_0)$.
Consider now
\begin{eqnarray*}\label{cons}
\dot{g}^{bf}=2H(\Sigma^b_a-\delta^b_a)g^{af}
\end{eqnarray*} 
Since the metric components are bounded the non-diagonal terms will contribute only with an $\epsilon$. Thus we have for every
component $g^{ij}$ (no summation over the indices in the following equation):
\begin{eqnarray*}
\dot{g}^{ij}=2H(\Sigma^i_i-1+\epsilon)g^{ij}\leq 2H (\max(\Sigma^i_i)-1+\epsilon)g^{ij}= 2H(-\frac34 +\epsilon)g^{ij}
\end{eqnarray*}
Using now the estimate of $H$
\begin{eqnarray}\label{dmetric}
\dot{g}^{ij}\leq t^{-1}(-1+\epsilon)g^{ij}
\end{eqnarray}
One can conclude that
\begin{eqnarray*}
 \| g^{-1}\| \leq O(t^{-1+\epsilon})
\end{eqnarray*}
From  (\ref{dmetric})
\begin{eqnarray*}
\dot{V}=\dot{g}^{bf}V_bV_f\leq t^{-1}(-1+\epsilon)V
\end{eqnarray*}
which means that
\begin{eqnarray*}
V=O(t^{-1+\epsilon})
\end{eqnarray*}
which gives us the same decay for $P$ as in the diagonal case:
\begin{eqnarray*}
P=O(t^{-\frac12+\epsilon})
\end{eqnarray*}
\subsubsection{Closing the bootstrap argument for Bianchi II}
It follows immediately by the same arguments as in the diagonal case:
\begin{eqnarray*}
\Sigma_-=O(t^{-1+\epsilon})\\
\Sigma_1^2=O(t^{-1+\epsilon})\\
\Sigma_1^3=O(t^{-1+\epsilon})\\
\Sigma_2^3=O(t^{-1+\epsilon})\\
\Sigma^2_3=O(t^{-1+\epsilon})
\end{eqnarray*}

Defining $(N_1)^2=-2N$ we arrive at:
\begin{eqnarray*}
&&\dot{\Sigma}_+=H[\frac{(N_1)^2}{3}-\Sigma_+(3+\frac{\dot{H}}{H^2})+\frac{4\pi}{3H^2}(3S^2_2+3S^3_3-2S)]\\
&&\dot{\Sigma}_-=H[-(3+\frac{\dot{H}}{H^2})\Sigma_-+\frac{4\pi(S^2_2-S^3_3)}{\sqrt{3}H^2}]\\
&&\dot{N}_1=H[(1+\frac{\dot{H}}{H^2}+4\Sigma_+)N_1+2\frac{W_{II}}{N_1}]
\end{eqnarray*}
Since $2\frac{W_{II}}{N_1}$ decays like $t^{-1+\epsilon}$ we see that we can apply
 the same arguments as in the diagonal case to obtain an improvement of the bootstrap assumptions:
\begin{eqnarray*}
 \Sigma_+-\frac18=O(t^{-\frac12+\epsilon})\\
\Sigma_-=O(t^{-\frac12+\epsilon})\\
N_1-\frac34=O(t^{-\frac12+\epsilon})
\end{eqnarray*}
The system which remains using the time variable $\tau$ is the following:
\begin{eqnarray*}
&&(\Sigma_2^1)'=\Sigma_2^1(q-2)+N^1_2-\frac{8\pi S^1_2}{H^2}\\
&&(\Sigma_3^1)'=\Sigma_3^1(q-2)+N^1_3-\frac{8\pi S^1_3}{H^2}\\
&& (N^1_2)'=-2(N_1)^2\Sigma^1_2-2(\Sigma_+-q-\sqrt{3}\Sigma_-)N^1_2+W^1_2\\
&& (N^1_3)'=-2(N_1)^2\Sigma^1_3-2(\Sigma_+-q+\sqrt{3}\Sigma_-)N^1_3+W^1_3
\end{eqnarray*}
Let us focus on the $\Sigma_2^1-N^1_2$-system. Using the estimates obtained we arrive at:
\begin{eqnarray*}
 \left(
\begin{matrix} \Sigma_2^1 \\
N^1_2
\end{matrix} \right)'=\left(
\begin{matrix} -\frac32 & 1  \\
-\frac98 & \frac34
\end{matrix} \right)\left(
\begin{matrix} \Sigma_2^1 \\
N^1_2
\end{matrix} \right)+O(\epsilon e^{(-\frac34 +\epsilon)\tau}) \left(
\begin{matrix} 1 \\
1
\end{matrix} \right)
\end{eqnarray*}
Let us go to the basis of eigenvectors of the linear system via the linear transformation
\begin{eqnarray*}
 \left(
\begin{matrix} \check{\Sigma}_2^1 \\
\check{N}^1_2
\end{matrix} \right)=\left(
\begin{matrix} \frac32 & -1  \\
-\frac32 & 2
\end{matrix} \right)\left(
\begin{matrix} \Sigma_2^1 \\
N^1_2
\end{matrix} \right)
\end{eqnarray*}
Thus we arrive at
\begin{eqnarray*}
 \left(
\begin{matrix} \check{\Sigma}_2^1 \\
\check{N}^1_2
\end{matrix} \right)'=\left(
\begin{matrix} -\frac34 & 0  \\
0 & 0
\end{matrix} \right)\left(
\begin{matrix} \check{\Sigma}_2^1 \\
\check{N}^1_2
\end{matrix} \right)+O(\epsilon e^{(-\frac34 +\epsilon)\tau}) \left(
\begin{matrix} 1 \\
1
\end{matrix} \right)
\end{eqnarray*}
Using the bootstrap assumptions for $\Sigma^1_2$ and $N^1_2$ we have an assumption for $\check{\Sigma}^1_2$. By the
usual contradiction argument we arrive at
\begin{eqnarray*}
\check{\Sigma}_2^1&=&\check{\Sigma}_2^1(\tau_0)e^{(-\frac34 +\epsilon)\tau}
\end{eqnarray*}
Integrating the equation for $\check{N}_2^1$ we arrive at
\begin{eqnarray*}
 \check{N}_2^1&=&\check{N}_2^1(\tau_0)+O(\epsilon)
\end{eqnarray*}
Going back to the variables $\Sigma_2^1$ and $N^1_2$ via
\begin{eqnarray*}
 \left(
\begin{matrix} \Sigma_2^1 \\
N^1_2
\end{matrix} \right)=\frac13\left(
\begin{matrix} 4 & 2  \\
3 & 3
\end{matrix} \right) \left(
\begin{matrix} \check{\Sigma}_2^1 \\
\check{N}^1_2
\end{matrix} \right)
\end{eqnarray*}
\begin{eqnarray*}
 \Sigma_2^1(\tau)&=& [2\Sigma_2^1(\tau_0)-\frac 43 N^1_2(\tau_0)]e^{(-\frac34 +\epsilon)\tau}+\frac43 N_2^1(\tau_0)-\Sigma^1_2(\tau_0)+O(\epsilon)\\
{N}_2^1(\tau)&=&[\frac32 {\Sigma}_2^1(\tau_0)-N^1_2(\tau_0)]e^{(-\frac34 +\epsilon)\tau}+2{N}_2^1(\tau_0)-\frac32 \Sigma_2^1(\tau_0)+O(\epsilon)
\end{eqnarray*}
Changing back to the time variable $t$:
\begin{eqnarray*}
 \Sigma_2^1(t)&=& C(t_0) [2\Sigma_2^1(t_0)-\frac 43 N^1_2(t_0)]t^{-\frac12 +\epsilon}+\frac43 N_2^1(t_0)-\Sigma^1_2(t_0)+O(\epsilon)\\
{N}_2^1(t)&=&C(t_0)[\frac32 {\Sigma}_2^1(t_0)-N^1_2(t_0)]t^{-\frac12 +\epsilon}+2{N}_2^1(t_0)-\frac32 \Sigma_2^1(t_0)+O(\epsilon)
\end{eqnarray*}
where $C$ is a constant, in particular $C(t_0)=t_0^{\frac12}e^{-\frac34 \tau_0}$. The only term which could prevent us from improving the estimates is the $\epsilon$ coming from the bootstrap assumptions of $\Sigma^1_2$, but note that
it comes in combination with $\Sigma^2_1$ as a product of both, thus the last term $O(\epsilon)$ on the right hand side of the last two equations does
not prevent us from improving our estimates. Thus if we wait long time enough and choose  
$N_2^1(t_0)$ and $\Sigma^1_2(t_0)$ small enough we will have an improvement for $N^1_2$ and $\Sigma^1_2$ since we can choose them 
independently and smaller then $A_{12}$ and $A_{c12}$.
There is no difference in the procedure for
$N^1_3$ and $\Sigma^1_3$.
\subsubsection{Arzela-Ascoli for Bianchi II}
Since all estimates have been improved we can apply Arzela-Ascoli and we arrive for $\Sigma^1_2$ and $N_2^1$ to:
\begin{eqnarray*}
 \Sigma_2^1(t=\infty)&=& \frac43 N_2^1(t_0)-\Sigma^1_2(t_0)\\
{N}_2^1(t=\infty)&=&2{N}_2^1(t_0)-\frac32 \Sigma_2^1(t_0)
\end{eqnarray*}
Consider now the following transformation of the basis vector
\begin{eqnarray*}
 &&\tilde{e}_1=e_1\\
&&\tilde{e}_2=e_2+ae_1\\
&&\tilde{e}_3=e_3+be_1
\end{eqnarray*}
It preserves the Lie-algebra, i.e. the Bianchi type. The following relation holds
between the variables $\Sigma^1_2$ and $\Sigma^1_3$ in the different basis:
\begin{eqnarray*}
\left(
\begin{matrix} \tilde{\Sigma}^1_1 & \tilde{\Sigma}^2_1 &\tilde{\Sigma}^3_1\\
\tilde{\Sigma}^1_2 & \tilde{\Sigma}^2_2 &\tilde{\Sigma}^3_2\\
\tilde{\Sigma}^1_3 & \tilde{\Sigma}^2_3 &\tilde{\Sigma}^3_3
\end{matrix} \right)=\left(\begin{matrix} 1 & 0 &0\\
a &1 &0\\
b & 0 &1\end{matrix} \right)\left(\begin{matrix}\Sigma^1_1 & 0 &0\\
\Sigma^1_2 &\Sigma^2_2 &0\\
\Sigma^1_3 & 0 &\Sigma^3_3
\end{matrix} \right)
\left(\begin{matrix} 1 & 0 &0\\
-a & 1 &0\\
-b & 0 &1\end{matrix} \right)=\left(\begin{matrix}\Sigma^1_1 & 0 &0\\
\Sigma^1_2+a(\Sigma^1_1-\Sigma_2^2) &\Sigma^2_2 &0\\
\Sigma^1_3+b(\Sigma^1_1-\Sigma_3^3) & 0 &\Sigma^3_3
\end{matrix} \right)\\
=\left(\begin{matrix}\Sigma^1_1 & 0 &0\\
\Sigma^1_2+a(3\Sigma_++\sqrt{3}\Sigma_-) &\Sigma^2_2 &0\\
\Sigma^1_3+b(3\Sigma_+-\sqrt{3}\Sigma_-) & 0 &\Sigma^3_3
\end{matrix} \right)
\end{eqnarray*}
We see that choosing $a=-\frac83 \Sigma^1_2(\infty)$ and $b=-\frac83 \Sigma^1_3(\infty)$ the transformed variables
 $\tilde{\Sigma}^1_2$, $\tilde{\Sigma}^1_3$ are zero asymptotically. By direct calculation one can
see that the same is true for the transformed variables $\tilde{N}^1_2$ and $\tilde{N}^1_3$. Thus we obtain
the same asymptotics as in the diagonal case and we can conclude:

\begin{thm}
Consider any $C^{\infty}$ solution of the Einstein-Vlasov system with Bianchi II symmetry and with $C^{\infty}$
 initial data. Assume that $|{\Sigma}_+(t_0)-\frac18|$, $|\Sigma_-(t_0)|$, $|\Sigma^1_2(t_0)|$, $|\Sigma^1_3(t_0)|$, $|\Sigma^2_3(t_0)|$,
 $|\Sigma^3_2(t_0)|$, $|\Sigma^2_1(t_0)|$, $|\Sigma^3_1(t_0)|$, $|N_1(t_0)-\frac{3}{4}|$, $|N_2^1(t_0)|$, $|N^1_3(t_0)|$  and $P(t_0)$ are sufficiently small. Then at
 late times, after possibly a basis change, the following estimates hold:
\begin{eqnarray*}
 H(t)&=&\frac{2}{3}t^{-1}(1+O(t^{-\frac{1}{2}}))\\
\Sigma_+-\frac{1}{8}&=&O(t^{-\frac{1}{2}})\\
\Sigma_-&=&O(t^{-1})\\
\Sigma^1_2&=&O(t^{-\frac{1}{2}})\\
\Sigma^1_3&=&O(t^{-\frac{1}{2}})\\
\Sigma^2_3&=&O(t^{-1})\\ 
\Sigma^3_2&=&O(t^{-1})\\ 
\Sigma^2_1&=&O(t^{-1})\\ 
\Sigma^3_1&=&O(t^{-1})\\ 
N_1-\frac{3}{4}&=&O(t^{-\frac{1}{2}})\\
N^1_2&=&O(t^{-\frac{1}{2}})\\
N^1_3&=&O(t^{-\frac{1}{2}})\\
P(t)&=&O(t^{-\frac{1}{2}})
\end{eqnarray*}
\end{thm}
\subsubsection{Closing the bootstrap argument of Bianchi VI$_0$}
It follows immediately by the same arguments as in the diagonal case:
\begin{eqnarray}
|\Sigma_2^1|=O(t^{-1+\epsilon})\\
|\Sigma_3^1|=O(t^{-1+\epsilon})\\
|\Sigma_2^3+\Sigma^2_3|=O(t^{-1+\epsilon})
\end{eqnarray}
Now consider the $\Sigma^3_2N_{23}$ system. Using the fact that $N^3_2=-N_{23}(N_3+N_2)$ we obtain
\begin{eqnarray*}
&&\dot{\Sigma}_2^3=H[-N_{23}(N_3+N_2)-\Sigma_2^3(3+\frac{\dot{H}}{H^2})-\frac{8\pi S^3_2}{H^2}]\\
&&\dot{N}_{23}=H[(2\Sigma_++2\sqrt{3}\Sigma_-+q)N_{23}+2\Sigma^3_2N_3-2\Sigma^1_2N_{13}]
\end{eqnarray*}
Using the bootstrap assumptions, the estimates obtained and the variable $\tau$:
\begin{eqnarray*}
 \left(
\begin{matrix} 
\Sigma^3_2\\
N_{23}
\end{matrix} \right)'=\left(
\begin{matrix} -\frac{3}{2}+\epsilon_1 & \epsilon_2 \\
-\frac{3}{2}+\epsilon_3 & \epsilon_1
\end{matrix} \right) \left(
\begin{matrix} 
\Sigma^3_2\\
N_{23}
\end{matrix} \right)+O(\epsilon e^{(-\frac32 +\epsilon)\tau}) \left(
\begin{matrix} 1 \\
1
\end{matrix} \right)
\end{eqnarray*}
where $\epsilon_1$, $\epsilon_2$ and $\epsilon_3$ have the following origin. The quantity $\epsilon_1$ is determined essentially by the error in $N$
and $\Sigma_+$ and note that $\Sigma^3_2$ comes in combination with $\Sigma^2_3$, thus this term can be chosen as small as we want. 
The quantity $\epsilon_2$ comes from $N_2+N_3$ and can be determined by the error of $N^2_2$ and finally the quantity $\epsilon_3$ which comes from
$N_3$ depends on the error of $N$, $N^2_2$ and $N_{23}^2$. Note in the last term that the quantity is squared, thus it is negligible. Having a look at 
the linearization and going to the eigenbasis via
\begin{eqnarray*}
 \left(
\begin{matrix} 
\check{\Sigma}^3_2\\
\check{N}_{23}
\end{matrix} \right)=\left(
\begin{matrix} 1 & 0 \\
-1 & 1
\end{matrix} \right) \left(
\begin{matrix} 
\Sigma^3_2\\
N_{23}
\end{matrix} \right)
\end{eqnarray*}
we come to the system
\begin{eqnarray*}
 \left(
\begin{matrix} 
\check{\Sigma}^3_2\\
\check{N}_{23}
\end{matrix} \right)'=\left(
\begin{matrix} -\frac32+\epsilon_1+\epsilon_2 & \epsilon_2 \\
\epsilon_3 - \epsilon_2 & \epsilon_1 - \epsilon_2
\end{matrix} \right) \left(
\begin{matrix} 
\check{\Sigma}^3_2\\
\check{N}_{23}
\end{matrix} \right)+O(\epsilon e^{(-\frac32 +\epsilon)\tau}) \left(
\begin{matrix} 1 \\
1
\end{matrix} \right)
\end{eqnarray*}
From which follows
\begin{eqnarray*}
&&\check{\Sigma}^3_2=\check{\Sigma}^3_2(\tau_0)e^{(-\frac32 +\epsilon)\tau}\\
&&\check{N}_{23}=\check{N}_{23}(\tau_0)+O(\epsilon)
\end{eqnarray*}
and going back 
\begin{eqnarray*}
&&\Sigma^3_2=\Sigma^3_2(\tau_0)e^{(-\frac32 +\epsilon)\tau}\\
&&N_{23}=N_{23}(\tau_0)-\Sigma^3_2(\tau_0)+O(\epsilon)
\end{eqnarray*}
We see that we have improved $N_{23}$, $\Sigma^3_2$ and with that also $\Sigma^2_3$

Using these estimates and the bootstrap assumptions let us focus now on the following system:
\begin{eqnarray*}
\dot{\Sigma}_+&=&H[\frac{N}{3}+\Sigma_+(q-2)+O(t^{-1+\epsilon})]\\
\dot{\Sigma}_-&=&H[-\frac{N^2_2}{\sqrt{3}}+(q-2)\Sigma_-+O(t^{-1+\epsilon})]\\
\dot{N}&=&H[2(2\Sigma_++q)N-4\sqrt{3}\Sigma_-N^2_2++O(t^{-1+\epsilon})]\\
\dot{N}^2_2&=&H[2(2\Sigma_++q)N^2_2+(\frac94\sqrt{3}+O(\epsilon))\Sigma_-+O(t^{-1+\epsilon})]
\end{eqnarray*}
where in the last equation $(N_2)^2+(N_3)^2$ was estimated with $N^2_2$, $N$ and $N_{23}$. The $O(\epsilon)$-term will not play a role since it can be
absorbed in the $\epsilon$ of the estimate. Let us look at the linearization using the variables $\tilde{\Sigma}_+=\Sigma_++\frac14$, $\tilde{\Sigma}_-=\Sigma_-$, $\tilde{N}=N+\frac98$, $\tilde{N}_2^2=N^2_2$ and
  and the variable $\tau$
\begin{eqnarray*}
 \left(
\begin{matrix} \tilde{\Sigma}_+ \\
\tilde{N} \\
\tilde{\Sigma}_-\\
\tilde{N^2_2}
\end{matrix} \right)'=\left(
\begin{matrix} -\frac{21}{16} & +\frac{5}{16} & 0& 0 \\
-\frac{45}{16} & -\frac{3}{16} & 0 & 0 \\
0 & 0 & -\frac{3}{2} &-\frac{\sqrt{3}}{3} \\
0 & 0 & \frac{9}{4}\sqrt{3} & 0
\end{matrix} \right)\left(
\begin{matrix} \tilde{\Sigma}_+ \\
\tilde{N} \\
\tilde{\Sigma}_-\\
\tilde{N^2_2}
\end{matrix} \right)
\end{eqnarray*}
The eigenvalues are 
\begin{eqnarray*}
 \lambda_{1/2}&=&-\frac{3}{4} \pm  \frac{3}{4}i\sqrt{3}\\
\lambda_{3/4}&=&-\frac{3}{4} \pm \frac{3}{4} i
\end{eqnarray*}
These eigenvalues are the same which appeared in the reflection symmetric case. Using the same arguments we arrive at
\begin{eqnarray*}
  |\Sigma_++\frac{1}{4}|\leq A_+ (1+t)^{-\frac{1}{2}+\epsilon}\\
|\Sigma_-|\leq A_- (1+t)^{-\frac{1}{2}+\epsilon}\\
|N+\frac{9}{8}|\leq A_{c1} (1+t)^{-\frac{1}{2}+\epsilon}\\
|N^2_2|\leq A_{c2} (1+t)^{-\frac{1}{2}+\epsilon}
\end{eqnarray*}
Finally, only $\Sigma^2_1(\tau_0)$, $\Sigma^3_1(\tau_0)$, $N_{12}(\tau_0)$ and $N_{13}(\tau_0)$ have to be improved. Let us look at the $\Sigma^2_1$,
$N_{12}$ system. There is no difference between this system and the $\Sigma^3_1$-$N_{13}$ system.
\begin{eqnarray*}
 \left(
\begin{matrix} 
\Sigma^2_1\\
N_{12}
\end{matrix} \right)'=\left(
\begin{matrix} -\frac{3}{2}+\epsilon_1 & \frac32+\epsilon_2 \\
-\frac{3}{2}+\epsilon_3 & \frac32+\epsilon_1
\end{matrix} \right) \left(
\begin{matrix} 
\Sigma^2_1\\
N_{12}
\end{matrix} \right)+O(\epsilon) \left(
\begin{matrix} 1 \\
1
\end{matrix} \right)
\end{eqnarray*}
\begin{eqnarray*}
 \lambda_1&=&\frac{1}{2} (2\epsilon_1-\sqrt{2}\sqrt{2\epsilon_2\epsilon_3-3\epsilon_2+3\epsilon_3})\\
\lambda_2&=&\frac{1}{2}  (2\epsilon_1+\sqrt{2}\sqrt{2\epsilon_2\epsilon_3-3\epsilon_2+3\epsilon_3})
\end{eqnarray*}
Now choosing the error of $N$ bigger than $\Sigma^{ab}\Sigma_{ab}$ $\epsilon_1$ will be negative. $\epsilon_2$ and $\epsilon_3$ can be chosen 
in such a way that the square root of the term is positive but in total smaller than $\epsilon_1$, such that we have two small and different eigenvalues.

\subsubsection{Arzela-Ascoli for Bianchi VI$_0$}
For Bianchi VI$_0$ we can apply Arzela-Ascoli as well. We see that $N_{23}$ will be zero and 
$\Sigma^2_1(\tau_0)$, $\Sigma^3_1(\tau_0)$, $N_{12}(\tau_0)$ and $N_{13}(\tau_0)$ will be come constants. This time we can make the following basis
change which preserves the Lie-algebra to obtain that the mentioned variables tend to zero:
\begin{eqnarray*}
 &&\tilde{e}_1=e_1+ae_2+be_3\\
&&\tilde{e}_2=e_2\\
&&\tilde{e}_3=e_3
\end{eqnarray*}
We can conclude
\begin{thm}
Consider any $C^{\infty}$ solution of the Einstein-Vlasov system with Bianchi VI$_0$ symmetry and with $C^{\infty}$
 initial data. Assume that $|{\Sigma}_+(t_0)+\frac14|$, $|\Sigma_-(t_0)|$, $\Sigma^1_2(t_0)$, $|\Sigma^1_3(t_0)|$, $|\Sigma^2_3(t_0)|$,
 $|\Sigma^3_2(t_0)|$, $|\Sigma^2_1(t_0)|$, $|\Sigma^3_1(t_0)|$, $|N(t_0)+\frac{9}{8}|$, $|N_2^2(t_0)|$,
 $|\Sigma^2_1(t_0)|$, $|N_{12}(t_0)|$,$|N_{13}(t_0)|$  and $P(t_0)$ are sufficiently small. Then at late times, after possibly a basis change, the following estimates hold:
\begin{eqnarray*}
 H(t)&=&\frac{2}{3}t^{-1}(1+O(t^{-\frac{1}{2}}))\\
\Sigma_+-\frac{1}{4}&=&O(t^{-\frac{1}{2}})\\
\Sigma_-&=&O(t^{-\frac12})\\
\Sigma^1_2&=&O(t^{-1})\\
\Sigma^1_3&=&O(t^{-1})\\
\Sigma^2_3&=&O(t^{-1})\\ 
\Sigma^3_2&=&O(t^{-1})\\ 
\Sigma^3_1&=&O(t^{-\frac12})\\ 
\Sigma^2_1&=&O(t^{-\frac12})\\
N_{12}&=&O(t^{-\frac12})\\
N_{13}&=&O(t^{-\frac12})\\
N_{23}&=&O(t^{-\frac12})\\
N_2^2&=&O(t^{-\frac12})\\
N+\frac98&=&O(t^{-\frac12})\\
P(t)&=&O(t^{-\frac{1}{2}})
\end{eqnarray*}
\end{thm}

\section{Conclusions}
As mentioned in the abstract the challenge here was to put the equations in a form such that the results of the diagonal case can be used. This can
be seen especially in the curvature variables. For Bianchi II it was sufficient to use the new variables $N^i_j=\frac{R^i_j}{H^2}$. 
For Bianchi VI$_0$ we had to introduce in addition to that the new variables $N_{ij}=\frac{g_{ij}}{g\sqrt{H}}$. The notation might be a little bit
confusing, but in both cases these variables have a connection to the curvature variables $N_1$, $N_2$ and $N_3$ of the diagonal case
 and this is the reason for the notation. In contrast to the diagonal case where the treatment of Bianchi II and VI$_0$ was almost identical, here
the latter case was more difficult. One reason could be the obvious increase in complexity. In the Bianchi II case it was sufficient to deal with 
$N$ instead of $N_1$ and look at the differences. In the case of Bianchi VI$_0$ $N$ had to be used to start the bootstrap argument. Then also $N^2_2$ 
\emph{and} $N_{23}$. This last variable made the correspondence to the diagonal case more difficult. As can been seen in the chapter where
the bootstrap argument was closed for Bianchi VI$_0$, we had to look more carefully on the dependence of the different $\epsilon$. Note also that we did not use
exactly the linearization in our last improvement of the estimates. We would have obtained that zero is an multiple eigenvalue and we would have not
obtained decay, but logarithmic growth. This would have been sufficient to close the bootstrap argument with corresponding suitable
bootstrap assumptions, but there would exist difficulties to apply the Arzela-Ascoli theorem and to obtain that the non-diagonal components 
become constant. Another difference to the diagonal case is the use of a basis change in the end. In general the non-diagonal components will become constants and thus
not relevant. However to obtain ``diagonal'' asymptotics a basis change will in general be necessary.

It would be interesting to investigate whether
the work on homogeneous Ricci solitons \cite{GIK} can help to understand the similarities and differences between
Bianchi II and VI$_0$ (in Thurstons classification Nil and Sol).

We have discussed the future asymptotics of some Bianchi models, what about the higher types? The case
of Bianchi VII$_0$ will probably be quite different. For instance in \cite{WHU} it was discovered that the Bianchi VII$_0$ spacetimes with a non-tilted fluid are not asymptotically self-similar in the future and
 that some oscillations take place. It is shown that dynamics are dominated by the Weyl curvature. However for dust a bifurcation of
 the Weyl curvature takes place (theorem 2.4 of \cite{WHU} and comments below). For this reason it is likely to expect difficulties
 when applying our techniques to this case. Something similar, but even more complicated happens in the case
 of Bianchi VIII spacetimes with a non-tilted fluid \cite{HHW}. 

What about inhomogeneous models? Some direction to generalize our results could be to analyze the Gowdy model which is the simplest
 inhomogeneous case. In \cite{RG} different links between Bianchi and (twisted) Gowdy spacetimes are considered, in particular for
 Bianchi I, II, VI$_0$ and VII$_0$. The analysis of perturbations is another
 interesting approach towards the understanding of inhomogeneous models (see \cite{Allen} for recent developments).

\end{document}